# From Collaboration to Solitude and Back: Remote Pair Programming during COVID-19

Darja Smite[1,2], Marius Mikalsen[2,3], Nils B. Moe[2,1], Viktoria Stray[2,4] and Eriks Klotins[1]

[1] Blekinge Institute of Technology, Karlskrona, Sweden
[2] SINTEF, Trondheim Norway
[3] Norwegian University of Science and Technology, Trondheim Norway
[4] University of Oslo, Oslo, Norway

`name.m.surname@[bth.se | sintef.no]`

**Abstract.** Along with the increasing popularity of agile software development, software work has become much more social than ever. Contemporary software teams rely on a variety of collaborative practices, such as pair programming, the topic of our study. Many agilists advocated the importance of collocation, face-to-face interaction, and physical artefacts incorporated in the shared workspace, which the COVID-19 pandemic made unavailable; most software companies around the world were forced to send their engineers to work from home. As software projects and teams overnight turned into distributed collaborations, we question what happened to the pair programming practice in the work-from-home mode. This paper reports on a longitudinal study of remote pair programming in two companies. We conducted 38 interviews with 30 engineers from Norway, Sweden, and the USA, and used the results of a survey in one of the case companies. Our study is unique as we collected the data longitudinally in April/May 2020, Sep/Oct 2020, and Jan/Feb 2021. We found that pair programming has decreased and some interviewees report not pairing at all for almost a full year. The experiences of those who paired vary from actively co-editing the code by using special tools to more passively co-reading and discussing the code and solutions by sharing the screen. Finally, we found that the interest in and the use of PP over time, since the first months of the forced work from home to early 2021, has admittedly increased, also as a social practice.

## 1    Introduction

Contemporary software engineering has become more social than ever [1]. The popularity of collaborative practices, such as pair programming (PP), have continuously grown along with the growing interest in implementing agile software development methodologies. The increased interest in PP has been linked with that joint problem solving outperforms individual capabilities [2] and that developers enjoy pairing more than working solo [3].

PP emerged as a collocated practice, sometimes even facilitated by specially dedicated work areas, i.e., open space with groups of workstations for PP [4], and relies on constant communication and collaboration. This is why, when the worldwide COVID-19 pandemic in 2020 has forced many software companies to send their employees to work from home (WFH), many of the collaborative practices that used to depend on physical collocation, were, obviously, disrupted. An interesting research question is then: **What happened to the PP practice in the forced WFH regime?**

We know that collaborative practices are more challenging in virtual teams [5]. The use of digital communication tools due to physical distance brings challenges, such as reduced communication quality due to poor network and meaning, tone, and emotion being lost and misunderstood over digital media [6]. Furthermore, facilitation of remote pair programming (RPP) requires specific tool support. Existing literature suggests that although several tools developed for RPP exist, there is very little empirical evaluation [7]. Even though distributed work is challenging, RPP has been shown to have benefits similar to collocated pairing [8]. However, there are not many studies on RPP in industry settings, and most are investigating students from a teaching perspective [7]. Further, there are no studies of teams that suddenly need to change from being collocated to working full-time remotely. All these research gaps motivated our study.

## 2  Background and Related Work

### 2.1  Pair programming

Pair programming (PP) is a key collaborative agile practice that is believed to improve team performance. In PP, two developers sit side-by-side at one computer, continuously collaborating on the same design, algorithm, code, or test [10]. In its original form, one developer takes a leading role (called the driver) while another (called the navigator) observes and actively provides feedback, asks questions and makes suggestions to ensure high quality of the produced code [11]. While the roles of the navigator and drivers are widely accepted, pairs often take on both responsibilities at the same time instead of having an explicit division of labor. Chong and Hurlbutt [12] found in their ethnographic study that having a strict separation of roles inhibits the natural way of working and that both developers having imminent access to the keyboard enabled rapid switching and made the developers more engaged. Further, Wray [13] reports additional scenarios of how two developers can collaborate on jointly improving the same code, for example, by jointly reviewing and discussing issues and potential solutions without any explicit roles.

The processes of PP relate to key processes of effective agile teamwork [9], which are monitoring, feedback, and backup behavior [14]. Since the pair might constantly change the driver and the navigator, or work without separating the roles, PP can help exercising backup behavior and establishing it in teams when missing [15]. PP is also found to be an efficient practice for education, and consequently a good practice when onboarding new people [2, 16]. Pairing involves shared decision making, therefore the practice supports self-management [17]. Further, recent research reports that PP is effective in terms of raising coding quality [16]. Last but not least, PP has also been found to be a practice that developers enjoy. Williams et al. [3] found that more than 90% stated that they enjoyed collaborative programming more than solo programming.

### 2.2 Remote pair programming

There are not many studies on remote pair programming (RPP) in industry, and most focus on students from a teaching perspective [7]. From existing studies we know that RPP provides benefits similar to colocated pairing such as increased productivity, code quality, and knowledge transfer in addition to the benefit of promoting communication between distributed team members [8]. At the same time, it is evident that RPP can be more challenging to initiate and perform compared to collocated PP. Initiation can be challenging because team members cannot just swivel their chairs around to the person sitting close to them. As such, it is important to have social software that can show who is available for pairing in distributed teams [5, 18]. In colocated PP, you work on the same code and frequently switch roles. In a distributed setting you need technology to support this [19]. Further, pairs also need to make voice and video calls and share screens during the coding activity. Because many tools exist, developers often have their own preferences on how to perform RPP, which sometimes leads to frustration within a team [20]. Therefore, having the right tools accepted by all team members is a success criterion for RPP to work, along with a list of important functions. The main requirements are [7, 19, 20] that pairs must be able to: 1) access, edit and synchronize the same files, 2) coordinate and fulfill the driver and observer roles, 3) point to different parts of the code, 4) know about the presence of their partner (text or video).

## 3 Empirical Cases and Research Method

### 3.1 The Case Study Design

In this paper, we report our findings from studying RPP forced by the COVID-19 pandemic in two companies. We chose a multiple longitudinal case study design [21] to understand what happens to distributed PP using digital collaboration tools. The case study is multiple, as we study engineers in two different companies. Our unit of study is individual engineers and their perceptions of and experiences with RPP. The study is longitudinal (see Fig. 1) as we have inquired the same people at different points of time. The cases are selected by convenience sampling, based on the availability of access to the company data and personnel and the interest of the companies in understanding the PP practice. The interviewees and survey respondents have volunteered to participate.

It is worth mentioning that we initially had access to a third company, in which PP was neither practiced systematically before transitioning to working from home, nor during the COVID-19 pandemic. Therefore, we decided not to include this case.

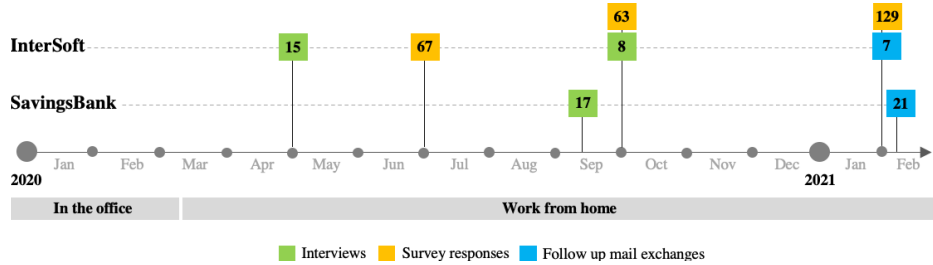

**Fig. 1.** Longitudinal data collection in each company

### 3.2 InterSoft and SavingsBank and Their Transition to WFH

The two companies we study are InterSoft and SavingsBank (both anonymized).

**InterSoft** is an international software company with development offices in Sweden, the UK, and the USA. In March 2020, InterSoft sent all employees in all locations to work from home, prohibiting access to the office spaces. The decision was initially set for two weeks with following extensions, which at the moment of our publication has reached September 2021. Thanks to the geographic distribution, the company has had the facilitating conditions, infrastructure and tools enabling distributed work before the pandemic. At the same time, InterSoft is an advanced agile company that promotes collaboration and teamwork, and agile practices, such as PP, are commonplace. Company culture seems to highly depend on intensive collaboration and collocation, this is why we were curious to study InterSoft's ability to transit into WFH. To support the employees, InterSoft has acquired various remote collaboration software licenses and launched a program for reimbursing office equipment in the early weeks of WFH and supported the transition to WFH through various experience sharing activities.

**SavingsBank** is a Norwegian software development company owned by an alliance of banks. In March 2020, its employees went from predominantly on-site work to 100% distributed work from home. Company sites were closed, and initially, employees were not allowed to be at the office at all. During the summer of 2020, because of a lower spread of the virus, the offices were open with restrictions (number of simultaneous employees at the office and in meeting rooms, and distance between employees). During the fall of 2020, the increased spread of the virus led to the offices being shut down again, which is still the case at the time of writing. Employees with a particular reason for being at the office (such as the need to run tests on a particular network or particularly problematic situation at home), are allowed to use the office. Notably, four of the 24 development teams in the company were partially distributed before, working across two different Norwegian cities, this is why SavingsBank also had facilitating conditions and infrastructure when moving into the WFH. The studied unit was described by practitioners as a leading agile environment in Norway, using state-of-the-art collaboration methods and technologies. Similar to Intersoft, SavingsBank had the technical infrastructure and tools that enable distributed work and a program for getting equipment from work. Through the year, employees were encouraged to keep their practices, and experiment with new practices and forms of digital collaboration.

### 3.3 Data Collection

This study is a part of a larger study of WFH experiences in each of the companies. PP emerged as one of the practices that changed, and thus became a candidate for a detailed analysis. Here, we report our findings from 38 interviews and 17 follow up inquiries from 30 engineers from the two companies, and a corporate tool satisfaction survey in one company. Our data collection is longitudinal, aiming at collecting experiences from same interviewees at different points of time to see changes in their experiences.

In InterSoft, we conducted 15 interviews in the first round of in April/May 2020. Interviewees were selected by convenience sampling, at the same time aiming at having representatives from the main locations (in Sweden and USA), age groups and family situations. The interviews were 45-60 min long and focused on the details of a typical day under WFH, reflections on the changes in the daily routines (schedule, tasks, meetings, teamwork, ceremonies, including PP), on what works and what does not work in the WFH mode, home office, and hopes for the future. We conducted follow up interviews with eight of the informants in September/October 2020. In the second round, the interviews were 30 min long and focused on the changes in routines and practices since the last inquiry. All interviews were semi-structured, conducted by two researchers in English via Zoom and audio-recorded with the consent of the interviewees. One of the interviewers led the interview, while the other took detailed notes (close to transcription). After the interview, notes were refined and complemented by cross-checking with the recording.

In SavingsBank, the interviews were conducted in September 2020. The interviewees were selected by tech leads to achieve a representative set of engineers. All interviews were 45-60 min long, conducted via MS Teams. Ten of the interviews were conducted by two researchers, and seven by one. The interviews were semi-structured, and focused on daily work practices using digital collaboration, interruptions, and internal open source. All interviews were audio-recorded with the consent of the interviewees and later transcribed. Notes were also taken during the interview. The interviews were conducted in Norwegian, the quotations from the interviews are thus translations.

For both companies we followed up interviewees with emails to aim at having multiple data points for each interviewee and to detail the RPP practices. The follow up inquiry was done in January/February 2021. We asked "Do you do RPP now?", "How often do you do RPP?", "Do you use any special tools for RPP?", "What do you think about RPP while working from home?". We received 21 answers (7 in InterSoft, and 14 in SavingsBank). As a result, we had four interviewees inquired (interviewed or emailed) three times, 22 inquired twice and only four inquired just once.

Additionally, we used results from an internal engineering satisfaction survey from InterSoft. We analyzed three survey rounds (Jul 2020, Oct 2020 and Jan 2021). We include responses regarding PP support: 67 responses from the first round, 63 responses from the second round and 129 responses from the third round. In the survey, respondents were asked to rate and comment on their satisfaction with the PP tools on a 5-point Likert scale (no support, poor support, moderate support, good support, great support). The survey was designed, ran, and analyzed independently from our study, qualifying it as secondary analysis of the secondary data [22].

### 3.4 Data Analysis

Our data analysis was conducted in several steps. First, we conducted interviews with a broad scope of inquiry to help companies understand how to cope with the new situation of working from home. After completing two rounds of interviews in InterSoft, and the first round of interviews in SavingsBank, the researchers involved in the interviews reflected on what emerged as the topics of interest for more detailed analysis.

Second, we presented preliminary findings to the companies and received their feedback. During the feedback session we took notes and later adjusted our interpretation of the findings based on the comments received.

Third, we analysed the written material, focusing on PP experiences. In the InterSoft case, we went through the written notes and partial transcripts from 23 interviews and found 21 references particularly relating to PP. In the SavingsBank case, the transcriptions were first coded in a bottom-up fashion, and 80+ codes were created, across 7 categories using the NVivo software for qualitative data analysis. Categories included concepts such as "new challenges during COVID". After this initial, close-to-the-text coding, we specifically looked for data on PP, and we found 15 references in the data.

Fourth, we started comparing the material for the two cases. Based on our initial understanding of the data, two of the researchers created an excel sheet in order to organize the data. The excel sheet had the following categories: "Whether or not PP is done", "How is PP done", "How often is PP done", "How often was PP done before working from home", and "Interesting quotes". The final category was used to capture interesting insights on the PP practice as experienced by the interviewees. These categories form the basis of the findings. In particular, we found the majority of the interviewees practiced PP before the pandemic, with the exception of few interviewees who did not; that the frequency of the use of RPP in WFH could vary between not at all, occasionally, regularly but less frequently than in the office, and more frequently than in the office; and for the way to perform RPP we ended up with two categories: Using special tools or Calling in & sharing screen. When the interview material was insufficient to answer the questions, we then sent follow up emails to inquire about more details. The feedback received was used to fill the gaps and enrich the findings.

Finally, the data from InterSoft was supplemented by the material from the tool satisfaction survey for triangulation and to seek explanations for the emerging findings.

### 3.5 Limitations and Threats to Validity

First, it's worth mentioning that our data collection initially was not focusing on the pair programming as the main topic of inquiry and the interview guides used in the two cases, although overlapping, were not fully aligned. We have mitigated this in the later follow-up inquiries, aligning the questions sent to the interviewees to mirror the data collected from the two cases.

Further, interviews as the data collection method have certain limitations. According to [21], interviews is one of the most important sources of case study information and should be considered "guided conversations rather than structured queries". And although interviews are insightful and provide perceived causal inferences and explanations, it is vital to be aware of the weaknesses of interviews as evidence. Interviews are likely to be biased, when it comes to poorly defined questions and responses. If the informant does not recall the past correctly, their answers are inaccurate, and they can be reflexive in the way that the informant gives what the interviewer wants to hear [21]. We sought to mitigate the shortcomings by collaboratively creating non-leading questions in the interview guide. Also, being semi-structured, we sought to follow up on the directions where the interviewee wanted to go. We also were, in most of the cases, two researchers doing the interviews, so we could adjust to each other, and in one of the companies the interviews were longitudinal, so we could check the same facts at two different points of time.

To further strengthen the validity of our findings, we performed triangulation by using data collected from different sources and by different methods. Triangulation is the core principle of case study research, which helps ensuring the consistency of the findings [21]. Besides, we presented our findings back to the companies, and sent the final version to all the interviewees to verify our analysis and interpretations.

There is also a limitation in terms of generalizing based on case studies. The main lesson learned in our work is the variety of experiences with RPP and the journey that individual engineers made, which can be viewed as a working hypothesis rather than strong theoretical claims [21]. What strengthens our findings is that we have two cases, spanning several locations. Finally, our findings are indicative as we shed empirical light on a relevant and emerging phenomenon, which could be relevant and interesting for individual engineers as well as software companies working with agile methods.

## 4      Findings

### 4.1     Do engineers pair program when working from home?

To understand the RPP as a practice, we asked the engineers whether or not they did PP when working from home. We grouped their responses into four categories; "Not all all", "Occasionally", "Regularly, but less frequently than in the office", and "Regularly and more frequently than in the office" (see Fig. 2).

Eight respondents state that they do not do RPP. Of these, two respondents state that it is because it was not done before either, and it is not something they have begun doing now. A third respondent comments that they became more separated during working from home, and that this makes it more challenging to do pair programming. An engineer that states that they do not do pair programming comments: "*We do little real pair programming. However, we try to make a habit out of discussing with the team when you start a task and discuss solutions. This is effective.*" (Oskar, Sep 2020)

Nine respondents state they are doing RPP, however very infrequent / occasionally, e.g., once a month. One respondent in this category is an engineer and a tech lead in a team. He describes doing a little bit of pair programming, but that he wants to do more. The challenges are related to the working situation at home and a lot of meetings: "*I have a home office together with my partner, and we switch between sitting in the kitchen and the bedroom when we have meetings at the same time. That makes it* *difficult to do it ad hoc. Additionally, I am one of those that has too many other meetings already, so it is hard to reserve time"* (Oliver, Sep 2020)

Another engineer, who used to do PP three times a week, notes how it is different doing PP while working from home *"That's not easy. There are a few tools to do it, like we have this Visual Code plugin – but it's just not the same thing, it's not as natural as it was in the office"* (Maya, Sep 2020). This echoes many others who said, that particularly in the beginning of working from home, using digital tools, like screen sharing or more customized tools like Tuple, was radically different from just walking over to the person next to you and sharing the keyboard and screen. Pair programming was more popular then, because it was easier, as one engineer comments: *"It was easier when you sat in the same location, [...] to take that little talk. If I was stuck, and knew that some other had solved the same problem last week, then it was much easier to just pop by and [ask] 'Can you quickly show me how you did it on your PC?', instead of starting to share screens, it is absolutely much better to do it when you are in the same location"* (Trond, Sep 2020)

In the interviews conducted in the early months of the forced WFH situation, many engineers confessed that they either have not done any PP or tried it only a few times. As an interviewee described: *"It's less easy to do [PP] with screen sharing. At least, I've been doing less of that. [...] It's been quite OK, but long term it would be good to have an easier possibility to do pair programming. But for these couple of months, it's OK"* (Ally, May 2020).

One of the reasons why PP failed to take off was that pairs needed to agree on doing PP and finding the time in a way that was not as natural as in the office. As one of the interviewees explained, in the early days of working from home, there were attempts to mimic the "old" way of working, where you could do more ad-hoc PP. But it obviously was not as easy: *"We've tried it once since we started working from home and I think it worked quite well, but I think it's hard, I am not sure why it is harder, but I've tried to say that I want to pair, and then it's Yeah, sure, let's do it after lunch. And then things change after lunch and then you don't do it. Maybe it's because we need to schedule this more explicitly. We have not figured out the solution"* (Sven, May 2020).

As a result, some teams started scheduling RPP sessions, as one explains: *"We started doing some more regular pair programming in our team recently. We booked a daily time slot for this, after our standup, and if during the standup some pairing opportunities come up, we use this dedicated time."* (Robert, Jan 2021)

A clear majority of responses at the end of our study period indicate that engineers pair program regularly. A more detailed analysis of the interviewee responses suggests that about half of those responding positively have increased their interest in pairing remotely over time. As people realised that remote work was not a temporary situation, the need for more collaborative practices increased and new ways of remote pairing were sought in order to address complex problems, like developing something new, or improving a design. The use of PP increased also because engineers became more experienced with remote work.

That both sentiments towards joint, digital collaboration practices, and the very remote practices themselves change is apparent from our data. Consider one engineer, who, when we interviewed him in Sep 2020, was very clear that the new, remote work was far inferior to being collocated. In Feb 2021, he describes how things have improved: *"Since the last time we talked, I think it has been a kind of development [...] My impression is that generally everyone became better at answering quickly and calls for handling issues which are difficult to [communicate in text]. I think everyone became better at using Slack and [MS] Teams correctly, and hence complex tasks are not considered that difficult anymore"* (Gustav, Feb 2021).

The progression of the interest in RPP over time is also supported by results of the tool support satisfaction survey from InterSoft conducted in Jul 2020 and repeated in Oct 2020 and Jan 2021 (See Fig. 3). The number of positive responses reporting good or great support increased from 54% to 73%.

Finally, two respondents stated that they do more PP than before the pandemic. Because of the need to plan and schedule collaboration, regular pairing sessions became a commodity. The members of the mentioned team do it directly after the standup and in the afternoon, on a daily (or almost daily) basis. One of the respondents explains how pair programming is a new initiative: "*We have in our area/team started focusing on pair programming now after the New year. We did it both with regard to quality, but not the least with regard to people's need for seeing each other and 'feel' that we work together while working from home.*" (Albert, Feb 2021)

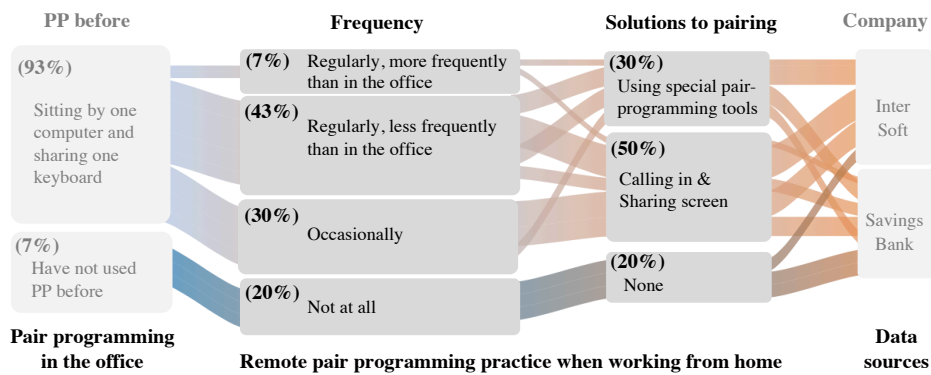

**Fig. 2.** An overview of PP in the studied companies: Transition from collocated PP or no PP to the way RPP is performed with regards to frequency and technical solutions.

### 4.2 How is remote pair programming done?

Similar changes as in the adoption of the RPP practice we also found also in the way it is performed. While in the early interviews conducted in April and May many reported to have not done any RPP or to have tried it once or twice, email responses received in January and February 2021 were full of detailed descriptions of tools that were tried, which were working well, and which setups were preferred. In our analysis of the different ways of doing RPP, we have derived two main categories of responses (see Solutions to pairing in Fig. 2).

In the first category, we have gathered experiences of RPP assisted by special tools that allow mimicking the collocated PP practice – looking at a code together and co-editing it and changing the driver and the navigator roles in the pair on a need basis: *"Pairing has gotten a lot better... We've gotten corporate licenses for some software that makes pairing easier. Not just screen sharing, but also controlling each other's computers, and being able to program simultaneously".* (Conor, Sep 2020)

Among the interviewees, we solicited experiences with Tuple, VSCode extension for PP and "Code with me" extension to IntelliJ. Evidently, the number of interviewees using special tools is not that large. We explain it with the lack of awareness of the tools, the initial skepticism towards the ability of the tools to support "real" PP and the difficulty to pair for a longer period of time.

The lack of awareness of the tools for RPP is also evident in the data collected through the tool satisfaction survey conducted at InterSoft (see Fig. 3). When asked how well the company tools support employees in performing PP tasks we see an increase in employees who are satisfied with the support. At the same time, fewer employees report poor or no support even in early 2021.

Our findings suggest an increasing adoption and support for RPP tools. The few employees discontent with the support suggest that the tools are not working for everyone. However, it is more likely that the respondents are not aware or have not tried out the tools available. The latter is also indirectly supported by our findings from the interviews regarding the frequency of use of RPP.

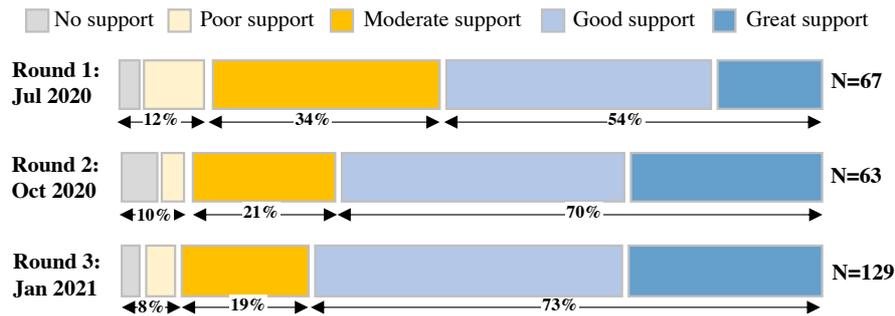

**Fig. 3.** Survey responses about satisfaction with tool support for pair programming

The next category reflects the adaptation of relatively passive PP practice. In fact, most of the interviewees in our study reported calling in their peers and looking at the code and discussing solutions while sharing their screen. Co-editing the code was not always a part of this practice. This is probably why, admittedly, this practice was not regarded as "real" PP, although still acknowledged to be better than working solo. In practice, one of the developers was said to occupy the driver's role, while the other is restricted to be the navigator, as switching is not technically possible. As one of the interviewees described it, *"[We] do not do it ´by the book´, but there is a lot of screen sharing together with one or more [team members]. This is related to debugging (checking logs/ config/ code), code assessment and implementing new functionality"* (Gustav, Feb 2021). Another interviewee explained, *"[It's] not like pair programming when you sit during four hours in parallel, but usually we do syncing for one hour, going through the code together, and then dividing the tasks, working for two hours in parallel and then syncing again. We call that 'more pair programming'."* (Ally, Sep 2020).

### 4.3 What are the main challenges in remote pair programming in WFH?

Pair programming "by the book" requires physical collocation, which for obvious reasons is impossible when everyone works from home. In the previous subsections, we described what tools and practices engineers follow to enable RPP, and mentioned that many experience difficulties and said that the fact that you need to do it in separation depending on the tools and internet is an obstacle to practicing PP. In the following we summarize the main challenges of remote setup according to our interviewees.

We learned that one of the key challenges associated with RPP is the very initiation of pairing. Several interviewees mentioned that the threshold to approach the colleagues was higher when working from home because they could not just shout out a question over the shoulder, as in the office, and did not have a good insight into what their colleagues are doing in a particular moment. As someone explained: *"You don't want to disturb people because you don't know what the others are doing right now"* (Ola, Apr 2020). This was even more challenging for the recently onboarded engineers who were not well familiar with other colleagues. As an interviewee explained: *"I prefer to do pair programming when I am at the office, because then it is easier to grab someone. I feel that there is a higher threshold when you are on Slack. The better you know your team, the easier it is."* (Kristine, Feb 2021)

Another important challenge was to find a technical solution that helps to work with the code together. RPP makes engineers depend on tools with varying quality and suitability for the task, for something that you previously did by walking over to the person next to you. Several interviewees said that tool-mediated PP would not feel the same, and therefore had not even tried it. Some tried to "jump on a call" but complained about the connection problems. Finally, some confessed that they wanted to use specific tools but they were not accepted as sufficiently secure in the company. Technical hiccups also contributed to the threshold, as an interviewee explained: *"It takes more time than usual to do [PP]. Difficult if someone gets technical issues with sharing the screen"* (Morten, Feb 2021). And another admitted: *"I do not enjoy PP while working from home, the software can be a bit slow at times, and it's easy to lose control of the cursor if both of us are trying to edit" (Carol, Feb 2021).*

Last but not least, several interviewees admitted that active tool-mediated PP was too intense and tiresome if done for hours, suggesting that a more passive mode was more suitable for the "online" version. As an interviewee explained: *"Proper pair programming is very intense and tiring, and it becomes difficult to hold for a longer time. But a more relaxed variant, where you have video and audio but work mostly separately and share a screen when needed, then you can work for hours"* (Gustav, Feb 2021).

### 4.4 What are the benefits of remote pair programming in WFH?

Despite these challenges, RPP still has a number of benefits. Pair programming is, in essence, a way for engineers to more quickly get a shared understanding of the problem, and then jointly and simultaneously work on solving the problem; means of seeking and receiving feedback and monitoring each other's work. The need for joint problem solving and acquiring a second opinion have not diminished when moving to working remotely, as one engineer notes: *"I feel like I have been an advocate for pairing up and pair programming and doing things together since I started, because I think that's a good way to avoid the buzz factor." (Kristoffer, May 2020)*

This notion is second by another engineer, who also points out how pair programming "*works well for doodling and problem solving" (Andreas, Feb 2021)*. In the formerly collocated contexts, engineers were used to working together, and, as a minimum, having a shared understanding of what they are going to do, before they split up and work individually to solve it.

We found that some respondents, despite admitting the difficulties of RPP and preferring to do it in co-location, acknowledge some benefits in tool-mediated pairing that are only available when using tools, such as the ability to control the partner's screen without taking away his or her keyboard. As an interviewee described: *"[Tuple] was the most recent tool that I've tried, and that was very good. [...] I think the big thing that Tuple has is the ability to type on your partner's screen. And I've used it twice recently and it was very quick and very easy, it was very helpful" (Conor, Sep 2020)*.

Additionally, there is a clear social component to RPP, which seems to become increasingly important as more time is spent away from the teammates. The following two quotes demonstrate how pairing enhances the work experience not only for the usual purposes, but for the sake of socialization: *"I think PP and writing code together is something we should do more of while working from home, both because two heads work better than one, but in particular because it is something social and it makes what you do feel more important" (Erik, Feb 2021)* and *"We did it both with regard to quality, but not the least with regard to people's need for seeing each other and to get a feel of working together although working from home [...] it adds something positive in terms of more contact with the other team members" (Albert, Sep 2020)*.

Another interesting observation relates to few interviewees who perceived RPP as easier because they did not disturb others unlike when working in the office. There was no longer the need to leave your desk and search for a quiet room to pair. As an interviewee explained: *"[Remote] pair programming works well and perhaps better than in the office where you would disturb the neighbour with the talking" (Arvid, Feb 2021)*.

Finally, we also associate RPP with more structured or disciplined daily routines. As one interviewee, who have admitted struggling with working from home in the beginning of the pandemic, described that frequent pairing helped him become more disciplined and keep the focus: *"While you are pairing, you don't get distracted, you don't want to get distracted. [...] The difference is, when I am soloing I usually sit in my living room, in a more casual setting, so I can relax, enjoy and have fun. But when I am pairing I feel like I should not be wasting their time, it should be more serious work. So I go to my home office with the chair and monitor to do it properly" (Ola, Sep 2020)*.

## 5   Discussion and Concluding Remarks

In this paper, we presented our findings from studying the changes in PP practices in two companies that sent their employees to work from home due to the COVID-19 outbreak. In the following, we summarize and discuss our findings.

In response to our RQ: "*What happened to PP practice in the forced WFH regime?*", we conclude that the overall use of PP has decreased, and some interviewees in our study report not pairing at all for almost a full year. However, we also found that the interest in and the use of RPP over time, since the first months of the forced WFH to early 2021, has admittedly increased, which is confirmed both by the interviewees and the results of the tool satisfaction survey in one of the companies. The experiences with RPP vary from actively co-editing the code by using special RPP tools (referred to as practicing PP "by the book") to co-reading and discussing the code and solutions by sharing screen (which supports the acceptance of alternative ways of pairing [13]). When it comes to the frequency of PP, we found that most do it less frequently than in the office. The sudden transition to the WFH led engineers to focus on individual tasks, temporarily reversing the social trend in software engineering [1, 2]. Our findings also illustrate how RPP is more intense and tiring than pairing in the office, supporting existing research on RPP being exhausting if performed for a long time [2]. Among the challenges, we found the difficulty of initiating RPP when merely mimicking the collocated practices. We can speculate that one reason for this is a lack of tools showing availability [5,18], or an initial failure to create new practices utilizing such tools. We also found how this changed over time. Our findings support the earlier research emphasizing the importance of and consensus over good tools for RPP [19, 20]. Concerning RPP benefits, our study confirms that engineers pair program because they enjoy it [3], especially as an important mean for socialization while WFH, which goes beyond RPP for quality and efficacy [11] and is not much discussed in the existing literature.

When reflecting on the individual journeys of engineers in our study made, we notice a few commonalities. We depict them in a conceptual representation of the acceptance of the work from home setup represented by an inverted Hype Curve[1] (see Fig. 4).

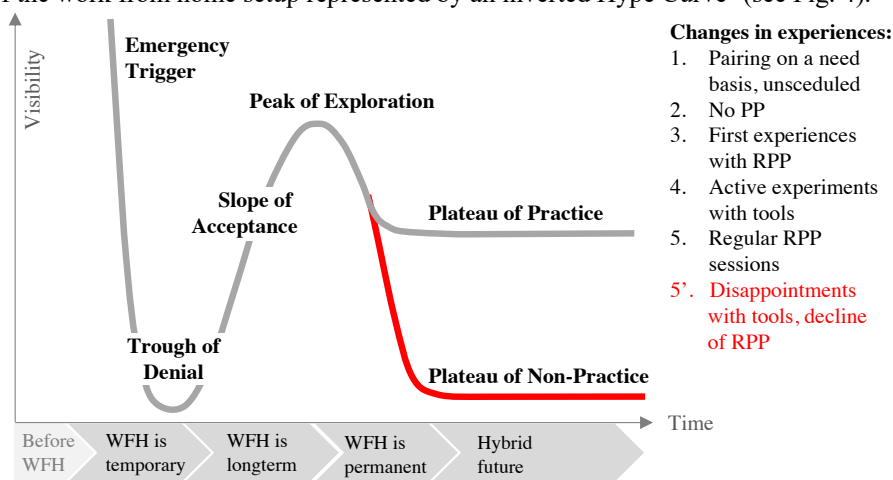

**Fig. 4.** Trends in acceptance of the WFH setup and respective changes in RPP experiences.

---

[1] Hype curve or cycle is a graphical representation of the maturity, adoption, and social application of technologies, which is developed, used and branded by the American research, advisory and information technology firm Gartner

14The traditional Hype Curve depicts the changes in the adoption of new technologies, starting with a technology breakthrough, the early publicity of which often inflates the expectations, shortly followed by the disillusionment, then continues with the work on advancing the technology, reflected in the enlightenment and increased popularity, finally leading to mainstream adoption. Unlike technology adoption, we believe that the transition to RPP resembles a vertically inverted curve. In the beginning, many of the interviewees did not do any RPP, because it was assumed to fail to provide the experience of the actual collocated PP as experienced in the office. As time went on and the perceiving temporary mode of working from home turned into a new way of living, engineers started experimenting with different ways of pairing. At the end of the first WFH year, RPP became more and more popular, not only as a way to solve complex problems or teach juniors, but also to facilitate collaboration and interaction and satisfy the socialization needs. We evidence the rise of experimentation with the tools and ways of approaching RPP (e.g., changes in the scheduling of regular pairing sessions). Evidently, not all interviewees have gone through the curve, indicating that the speed with the RPP adoption varies.

We can expect those who, at the moment of our study, have followed remote pairing occasionally, or who have not yet found the proper tool support, will do so soon. Our hopes are supported by stories we have collected from those who indicate an increasing interest in remote pairing over time.

Yet, we also found that several engineers who used to pair program in the office, have not done it since they were forced to work from home. Besides, our results suggest that remote collaboration is not as natural as the collaboration in the office, and that the success of remote collaboration reported by the interviewees in our study highly depends on the existing social connections. If the future will be a hybrid of remote and office work, and companies will more willingly hire experts from remote locations, we will no longer be able to rely or assume that people will know each other well. This leads to a question, whether future teams will find ways to overcome the threshold of initiating and maintaining the high level of collaboration or regress to a more transaction-relationship. From one hand, PP can become a practice to familiarize the team members and keep the collaboration high even when team members choose to continue working remotely. On the other hand, if the perception of being an "unnatural" or "challenging" practice will dominate the mainstream opinion, PP might become extinct in the repertoire of the future software teams.

We believe that given many of the positive effects reported from working from home, findings ways to keep PP in the repertoire of the future teams are important. The understanding of the RPP and the very nature of future collaborations at the virtual workplace is one important direction for future research.

## Acknowledgements

This research is funded by the Swedish Knowledge Foundation within the ScaleWise project (KK-Hög grant 2019/0087) and the S.E.R.T. research profile project, and the Research Council of Norway through the 10xTeams project (grant 309344).